# Spectral decomposition approach to macroscopic parameters of Fokker-Planck flows: Part 1

*Igor A. Tanski*
*Moscow, Russia*
*tanski.igor.arxiv@gmail.com*

*ABSTRACT*

In this paper we proceed with investigation of connections between Fokker - Planck equation and continuum mechanics. In spectral decomposition of Fokker - Planck equation solution we preserve only terms with the smallest degree of damping. We find, that macroscopic parameters of Fokker-Planck flows, obtained in this way, have following properties: velocities field possess potential, its potential is proportional to density logarithm and satisfy diffusion equation. We proved, that such a pair of density and velocities field satisfy the set of classic hydrodynamics equations for isothermal compressible fluid with friction mass force, proportional to velocity. We proved also, that the potential velocities field alone, with potential, which satisfy diffusion equation, satisfy Burgers equation without mass forces.

In this paper we proceed with investigation of connections between Fokker - Planck equation and continuum mechanics. In our work [1] we find, that by certain assumptions (Maxwellian distributions for velocities and big enough time and small deviations of stresses from hydrostatic pressure) we can deduct equations of classic hydrodynamics from Fokker-Planck equation. Between these assumptions Maxwellian distributions for velocities is most questionable. In this paper we try to get results in more direct way. We follow the method, proposed in [3]. Namely, in spectral decomposition of solution we preserve only terms with the smallest degree of damping. In the course of calculations we use results of our work [2].

In the next part we shall discuss results of the next degree terms consideration.

# 1. Introduction

In this section we remind main results of our work [2].

The expression for space density of continuum is

$$\rho = \int\limits_{-\infty}^{+\infty}\int\limits_{-\infty}^{+\infty}\int\limits_{-\infty}^{+\infty} \sum_{p_1=0}^{\infty}\sum_{p_2=0}^{\infty}\sum_{p_3=0}^{\infty} \exp\left[-t\left(\alpha\sum_{j=1}^{3}p_j + \frac{k}{\alpha^2}\sum_{j=1}^{3}\omega_j^2\right)\right]\times \quad (1)$$

$$\times A_{\omega_1\omega_2\omega_3 p_1 p_2 p_3}\left(\frac{k}{2\pi\alpha}\right)^{3/2}\prod_{j=1}^{j=3}(-i)^{p_j}\left(\frac{2k}{\alpha}\right)^{p_j/2}\exp(i\omega_j x_j)\exp\left[-\frac{k}{2\alpha}\left(\frac{\omega_j}{\alpha}\right)^2\right]\left(-\frac{\omega_j}{\alpha}\right)^{p_j}d\omega_1 d\omega_2 d\omega_3.$$

The expression for average velocities is

$$u_k = \frac{i}{\rho}\int\limits_{-\infty}^{+\infty}\int\limits_{-\infty}^{+\infty}\int\limits_{-\infty}^{+\infty} \sum_{p_1=0}^{\infty}\sum_{p_2=0}^{\infty}\sum_{p_3=0}^{\infty} \exp\left[-t\left(\alpha\sum_{j=1}^{3}p_j + \frac{k}{\alpha^2}\sum_{j=1}^{3}\omega_j^2\right)\right]\times \quad (2)$$

$$\times A_{\omega_1\omega_2\omega_3 p_1 p_2 p_3}\left(\frac{k}{2\pi\alpha}\right)^{3/2}\left[-\frac{k}{\alpha}\frac{\omega_k}{\alpha} - \frac{\alpha p_k}{\omega_k}\right]\times$$

$$\times \prod_{j=1}^{j=3}(-i)^{p_j}\left(\frac{2k}{\alpha}\right)^{p_j/2}\exp(i\omega_j x_j)\exp\left[-\frac{k}{2\alpha}\left(\frac{\omega_j}{\alpha}\right)^2\right]\left(-\frac{\omega_j}{\alpha}\right)^{p_j}d\omega_1 d\omega_2 d\omega_3.$$

The expression for current of momentum tensor is

$$J_{kl} = -\int\limits_{-\infty}^{+\infty}\int\limits_{-\infty}^{+\infty}\int\limits_{-\infty}^{+\infty} \sum_{p_1=0}^{\infty}\sum_{p_2=0}^{\infty}\sum_{p_3=0}^{\infty} \exp\left[-t\left(\alpha\sum_{j=1}^{3}p_j + \frac{k}{\alpha^2}\sum_{j=1}^{3}\omega_j^2\right)\right]\times \quad (3)$$

$$\times A_{\omega_1\omega_2\omega_3 p_1 p_2 p_3}\left(\frac{k}{2\pi\alpha}\right)^{3/2}\left(\left[-\frac{k}{\alpha}\frac{\omega_k}{\alpha} - \frac{\alpha p_k}{\omega_k}\right]\left[-\frac{k}{\alpha}\frac{\omega_l}{\alpha} - \frac{\alpha p_l}{\omega_l}\right] - \delta_{kl}\left[\frac{k}{\alpha} + \frac{\alpha^2 p_l}{\omega_l^2}\right]\right)\times$$

$$\times \prod_{j=1}^{j=3}(-i)^{p_j}\left(\frac{2k}{\alpha}\right)^{p_j/2}\exp(i\omega_j x_j)\exp\left[-\frac{k}{2\alpha}\left(\frac{\omega_j}{\alpha}\right)^2\right]\left(-\frac{\omega_j}{\alpha}\right)^{p_j}d\omega_1 d\omega_2 d\omega_3.$$

Stresses are equal to

$$\sigma_{kl} = \rho u_k u_l - J_{kl}. \quad (4)$$

Hydrostatic pressure is

$$p = -\frac{1}{3}\sigma_{kk}. \quad (5)$$

We use usual summation convention: indices that occur twice are considered to be summed over.

Coefficients $A_{\omega_1\omega_2\omega_3 p_1 p_2 p_3}$, present in (1-5), are equal to:

$$A_{\omega_1\omega_2\omega_3 p_1 p_2 p_3} = \frac{1}{2^{p_1+p_2+p_3} p_1! p_2! p_3!} \left(\frac{\alpha}{2\pi k}\right)^{\frac{3}{2}} \frac{1}{(2\pi)^3} \exp\left[\frac{2k}{\alpha^3}\left(\omega_1^2 + \omega_2^2 + \omega_3^2\right)\right] \times \qquad (6)$$

$$\times \int_{-\infty}^{\infty} dx_1 \int_{-\infty}^{\infty} dx_2 \int_{-\infty}^{\infty} dx_3 \int_{-\infty}^{\infty}\int_{-\infty}^{\infty}\int_{-\infty}^{\infty} n_0(x_j, v_j)\ \psi_{\omega_1\omega_2\omega_3 p_1 p_2 p_3} dv_1 dv_2 dv_3.$$

where $n_0(x_j, v_j)$ - initial (at $t = 0$) value of Fokker - Planck equation independent variable $n$ ;
$\psi_{\omega_1\omega_2\omega_3 p_1 p_2 p_3}$ - eigenfunction of adjoint differential operator with eigenvalues $\omega_1\omega_2\omega_3 p_1 p_2 p_3$.

Expression for eigenfunctions of adjoint differential operator is

$$\psi_{\omega_1\omega_2\omega_3 p_1 p_2 p_3} = \prod_{j=1}^{j=3} \exp\left(-i\omega_j(x_j + \frac{v_j}{\alpha})\right) H_{p_j}\left(\sqrt{\frac{\alpha}{2k}}\left(v_j + \frac{2i\omega_j k}{\alpha^2}\right)\right) \qquad (7)$$

## 2. Expressions for macroscopic parameters

We preserve the terms with biggest possible eigenvalue for each space mode. For this purpose we make substitution $p_j = 0$ for $j = 1, 2, 3$. This means, that for each space mode $\omega_j$ we select velocity distribution with slowest damping.

In this way we get following expression for density

$$\rho = \int_{-\infty}^{+\infty}\int_{-\infty}^{+\infty}\int_{-\infty}^{+\infty} \exp\left[-t\ \frac{k}{\alpha^2} \sum_{j=1}^{3} \omega_j^2\right] \times \qquad (8)$$

$$\times A_{\omega_1\omega_2\omega_3 000} \left(\frac{k}{2\pi\alpha}\right)^{3/2} \prod_{j=1}^{j=3} \exp(i\omega_j x_j) \exp\left[-\frac{k}{2\alpha}\left(\frac{\omega_j}{\alpha}\right)^2\right] d\omega_1 d\omega_2 d\omega_3.$$

We see from (8), that $\rho$ satisfy diffusion equation

$$\frac{\partial \rho}{\partial t} = \frac{k}{\alpha^2}\left(\frac{\partial^2 \rho}{\partial x^2} + \frac{\partial^2 \rho}{\partial y^2} + \frac{\partial^2 \rho}{\partial z^2}\right). \qquad (9)$$

In the same way we obtain expression for velocities

$$u_k = \frac{i}{\rho} \int_{-\infty}^{+\infty}\int_{-\infty}^{+\infty}\int_{-\infty}^{+\infty} \exp\left[-t\ \frac{k}{\alpha^2} \sum_{j=1}^{3} \omega_j^2\right] \times \qquad (10)$$

$$\times A_{\omega_1\omega_2\omega_3 000} \left(\frac{k}{2\pi\alpha}\right)^{3/2} \left[-\frac{k}{\alpha}\frac{\omega_k}{\alpha}\right] \prod_{j=1}^{j=3} \exp(i\omega_j x_j) \exp\left[-\frac{k}{2\alpha}\left(\frac{\omega_j}{\alpha}\right)^2\right] d\omega_1 d\omega_2 d\omega_3,$$

or

$$\rho u_j = -\frac{k}{\alpha^2}\frac{\partial \rho}{\partial x_j}. \tag{11}$$

Note, that though we dropped all fast damping terms, (8) and (10) are still derived from solution of Fokker - Planck equation. Therefore they must satisfy certain conservation laws.

The mass conservation law

$$\frac{\partial \rho}{\partial t} + \frac{\partial}{\partial x_j}\left(\rho u_j\right) = 0. \tag{12}$$

we deduct easily from (9) and (11).

The expression for current of momentum tensor is

$$J_{kl} = -\int_{-\infty}^{+\infty}\int_{-\infty}^{+\infty}\int_{-\infty}^{+\infty} \exp\left[-t\frac{k}{\alpha^2}\sum_{j=1}^{3}\omega_j^2\right] A_{\omega_1\omega_2\omega_3 000}\left(\frac{k}{2\pi\alpha}\right)^{3/2}\left[\left(\frac{k}{\alpha^2}\right)^2 \omega_k \omega_l - \frac{k}{\alpha}\delta_{kl}\right]\times \tag{13}$$

$$\times \prod_{j=1}^{j=3}\exp(i\omega_j x_j)\exp\left[-\frac{k}{2\alpha}\left(\frac{\omega_j}{\alpha}\right)^2\right] d\omega_1 d\omega_2 d\omega_3,$$

or

$$J_{ij} = \left(\frac{k}{\alpha^2}\right)^2 \frac{\partial^2 \rho}{\partial x_i \partial x_j} + \frac{k}{\alpha}\rho\delta_{ij}. \tag{14}$$

$$\frac{\partial \rho}{\partial x_j} = -\frac{\alpha^2}{k}\rho u_j. \tag{15}$$

We can express $\rho$ space derivatives in terms of velocities and their space derivatives in the following way :

$$\frac{\partial^2 \rho}{\partial x_i \partial x_j} = -\frac{\alpha^2}{k}\rho\frac{\partial u_i}{\partial x_j} + \left(\frac{\alpha^2}{k}\right)^2 \rho u_i u_j; \tag{16}$$

It follows, that

$$\frac{\partial u_i}{\partial x_j} - \frac{\partial u_j}{\partial x_i} = 0, \tag{17}$$

that is velocities field possess potential. From (11) we see, that this potential is proportional to $\ln(\rho)$.

$$\frac{\partial^2 \rho}{\partial x_i \partial x_j} = -\frac{\alpha^2}{k} \rho \frac{1}{2}\left(\frac{\partial u_i}{\partial x_j} + \frac{\partial u_j}{\partial x_i}\right) + \left(\frac{\alpha^2}{k}\right)^2 \rho u_i u_j. \tag{18}$$

The final expression for current of momentum tensor is

$$J_{ij} = -\frac{k}{\alpha^2} \rho \frac{1}{2}\left(\frac{\partial u_i}{\partial x_j} + \frac{\partial u_j}{\partial x_i}\right) + \rho u_i u_j + \frac{k}{\alpha} \rho \delta_{ij}. \tag{19}$$

Tensor of stresses components are equal to

$$\sigma_{ij} = \rho u_i u_j - J_{ij} = \frac{k}{\alpha^2} \rho \frac{1}{2}\left(\frac{\partial u_i}{\partial x_j} + \frac{\partial u_j}{\partial x_i}\right) - \frac{k}{\alpha} \rho \delta_{ij}. \tag{20}$$

We get once again (compare with eq. (41) [1]) the state equation of compressible viscous fluid. Kinematic viscosity is equal to $\nu = \dfrac{k}{\alpha^2}$.

Hydrostatic pressure is

$$p = -\frac{1}{3}\sigma_{kk} = \frac{k}{\alpha}\rho - \frac{1}{3}\frac{k}{\alpha^2}\rho \frac{\partial u_q}{\partial x_q}. \tag{21}$$

Relation between density and hydrostatic pressure is $p = (k/\alpha)\rho$. It coincides with ideal gas isotherm. This is natural, because Fokker - Planck continuum is in thermal equilibrium with enclosing motionless medium.

## 3. Equations of movement

Now we are prepared to consider another conservation law - equation of movement (which in our case is Navier - Stokes equation). The equation is:

$$\frac{\partial(\rho u_i)}{\partial t} + \frac{\partial(\rho u_i u_j)}{\partial x_j} - \frac{\partial \sigma_{ij}}{\partial x_j} + \alpha \rho u_i = 0. \tag{22}$$

Use (11) and substitute (20)

$$-\frac{k}{\alpha^2}\frac{\partial^2 \rho}{\partial x_i \partial t} + \frac{\partial}{\partial x_j}\left(\rho u_i u_j - \frac{k}{\alpha^2}\rho\frac{1}{2}\left(\frac{\partial u_i}{\partial x_j} + \frac{\partial u_j}{\partial x_i}\right) + \frac{k}{\alpha}\rho\delta_{ij}\right) + \alpha \rho u_i = 0. \tag{23}$$

Use first conservation law (12) and drop according to (15) 2 last terms.

$$\frac{k}{\alpha^2} \frac{\partial^2}{\partial x_i \partial x_j}\left(\rho u_j\right) + \frac{\partial}{\partial x_j}\left(-\frac{k}{\alpha^2} \rho \frac{1}{2}\left(\frac{\partial u_i}{\partial x_j} + \frac{\partial u_j}{\partial x_i}\right) + \rho u_i u_j\right) = 0, \quad (24)$$

or

$$\frac{\partial}{\partial x_j}\left(\frac{k}{\alpha^2} \frac{\partial}{\partial x_i}\left(\rho u_j\right) - \frac{k}{\alpha^2} \rho \frac{1}{2}\left(\frac{\partial u_i}{\partial x_j} + \frac{\partial u_j}{\partial x_i}\right) + \rho u_i u_j\right) = 0. \quad (25)$$

Now use (15) to simplify the first term and get

$$\frac{\partial}{\partial x_j}\left(\frac{k}{\alpha^2} \rho \frac{\partial u_j}{\partial x_i} - \frac{k}{\alpha^2} \rho \frac{1}{2}\left(\frac{\partial u_i}{\partial x_j} + \frac{\partial u_j}{\partial x_i}\right)\right) = 0. \quad (26)$$

This is true according to (17).

In this way we checked, that variables $\rho$ and $v_i$ satisfy the set of equations of classic hydrodynamics of isothermal compressible fluid.

We proved, that the pair of potential velocities field and its potential, which must satisfy diffusion equation, satisfy the set of classic hydrodynamics equations for isothermal compressible fluid with friction mass force, proportional to velocity.

## 4. Burgers equation

We can write equations of movement in slightly modified form

$$\frac{\partial u_i}{\partial t} + u_j \frac{\partial u_i}{\partial x_j} - \frac{1}{\rho}\frac{\partial \sigma_{ij}}{\partial x_j} + \alpha u_i = 0. \quad (27)$$

Substitute expression for stresses (20)

$$\frac{\partial u_i}{\partial t} + u_j \frac{\partial u_i}{\partial x_j} - \frac{1}{\rho}\frac{\partial}{\partial x_j}\left[\frac{k}{\alpha^2} \rho \frac{1}{2}\left(\frac{\partial u_i}{\partial x_j} + \frac{\partial u_j}{\partial x_i}\right) - \frac{k}{\alpha}\rho\delta_{ij}\right] + \alpha u_i = 0. \quad (28)$$

Drop two last terms according to (15) and open the brackets. We exclude $\rho$ and all its space derivatives

$$\frac{\partial u_i}{\partial t} + u_j \frac{\partial u_i}{\partial x_j} - \frac{k}{2\alpha^2}\frac{\partial^2 u_i}{\partial x_j \partial x_j} - \frac{k}{2\alpha^2}\frac{\partial^2 u_j}{\partial x_i \partial x_j} + u_j \frac{1}{2}\left(\frac{\partial u_i}{\partial x_j} + \frac{\partial u_j}{\partial x_i}\right) = 0. \quad (29)$$

As a result we get following equation for $u_i$ only. This equation strongly resembles Burgers equation.

$$\frac{\partial u_i}{\partial t} + 2u_j \frac{\partial u_i}{\partial x_j} - \frac{k}{\alpha^2} \frac{\partial^2 u_i}{\partial x_j \partial x_j} = 0, \tag{30}$$

To get more usual form of Burgers equation we perform substitution $t' = 2t$ and $v' = \frac{k}{2\alpha^2}$.

$$\frac{\partial u_i}{\partial t'} + u_j \frac{\partial u_i}{\partial x_j} - v' \frac{\partial^2 u_i}{\partial x_j \partial x_j} = 0. \tag{31}$$

This is well known Burgers equation. Our expressions (9) and (15) build the Hopf - Cole transformation [4].

We proved, that potential velocities field, with potential, which satisfy diffusion equation, satisfy Burgers equation.

**DISCUSSION**

In this paper we proceed with investigation of connections between Fokker - Planck equation and continuum mechanics. We base upon expressions from our work [2]. In spectral decomposition of Fokker - Planck equation solution we preserve only terms with the smallest degree of damping. We find, that macroscopic parameters of Fokker-Planck flows, obtained in this way, have following properties: velocities field possess potential, its potential is proportional to density logarithm and satisfy diffusion equation. This property of density is well known (see [3]), the property of velocity field seems new. We proved, that such a pair of density and velocities field satisfy the set of classic hydrodynamics equations for isothermal compressible fluid with friction mass force, proportional to velocity. We proved also, that the potential velocities field alone, with potential, which satisfy diffusion equation, satisfy Burgers equation without mass forces. This is evident from theory of Hopf - Cole transformation (see [4]), but in connection with other results provides interesting example of exact linearization using spectral decomposition (see [2]). In the next part of this article we shall examine results of the next degree terms consideration.

---